\documentstyle[11pt]{article}

\begin{document}
\large

\def\lsim{\mathrel{\rlap{\lower3pt\hbox{\hskip0pt$\sim$}}
    \raise1pt\hbox{$<$}}}         
\def\gsim{\mathrel{\rlap{\lower4pt\hbox{\hskip1pt$\sim$}}
    \raise1pt\hbox{$>$}}}         
\def\dblint{\mathop{\rlap{\hbox{$\displaystyle\!\int\!\!\!\!\!\int$}}
    \hbox{$\bigcirc$}}}
\def\ut#1{$\underline{\smash{\vphantom{y}\hbox{#1}}}$}

\newcommand{\beq}{\begin{equation}}
\newcommand{\eeq}{\end{equation}}
\newcommand{\dem}{\Delta M_{\mbox{B-M}}}
\newcommand{\dega}{\Delta \Gamma_{\mbox{B-M}}}

\newcommand{\ind}[1]{_{\begin{small}\mbox{#1}\end{small}}}
\newcommand{\WA}{{\em WA}}
\newcommand{\SM}{Standard Model }
\newcommand{\QCD}{{\em QCD }}
\newcommand{\KM}{{\em KM }}
\newcommand{\hscale}{\mu\ind{hadr}}
\newcommand{\aver}[1]{\langle #1\rangle}

\newcommand{\appa}{\mbox{\ae}}
\newcommand{\CP}{{\em CP}}
\newcommand{\fy}{\varphi}
\newcommand{\hi}{\chi}
\newcommand{\al}{\alpha}
\newcommand{\as}{\alpha_s}
\newcommand{\gf}{\gamma_5}
\newcommand{\gm}{\gamma_\mu}
\newcommand{\gn}{\gamma_\nu}
\newcommand{\be}{\beta}
\newcommand{\ga}{\gamma}
\newcommand{\de}{\delta}
\renewcommand{\Im}{\mbox{Im}}
\renewcommand{\Re}{\mbox{Re}}
\newcommand{\GeV}{\,\mbox{GeV }}
\newcommand{\MeV}{\,\mbox{MeV }}
\newcommand{\matel}[3]{\langle #1|#2|#3\rangle}
\newcommand{\state}[1]{|#1\rangle}
\newcommand{\ra}{\rightarrow}
\newcommand{\ve}[1]{\vec{\bf #1}}

\vspace*{.4cm}
\begin{flushright}
\large{
CERN-TH.7020/93\\
UND-HEP-93-BIG\hspace*{0.1em}02}\\
September 1993\\
\end{flushright}
\vspace{0.4cm}
\begin{center} \LARGE {
Weak Annihilation and the End Point
Spectrum in Semileptonic B Decays}
\end{center}
\vspace{.4cm}
\begin{center} \Large I.I.Bigi
\\
{\normalsize{\it Theoretical Physics Division, CERN\\
CH-1211 Geneva 23, Switzerland}\\
and\\
{\it Dept.of Physics,
University of Notre Dame du
Lac, Notre Dame, IN 46556, U.S.A.
\footnote{permanent address}
\\e-mail address: VXCERN::IBIGI, BIGI@UNDHEP}}

\vspace{.4cm}
N.G.Uraltsev
\\{\normalsize\it Dept.of
Physics,
University of Notre Dame du Lac, Notre Dame, IN 46556
\\{\normalsize and}\\
St.Petersburg Nuclear Physics Institute,
Gatchina, St.Petersburg 188350, Russia $^1$
\\e-mail address: URALTSEV@UNDHEP, URALTSEV@LNPI.SPB.SU}
\end{center}
\thispagestyle{empty} \vspace{.4cm}

\centerline{\Large\bf Abstract}
\vspace{.4cm}

A general relationship is formulated between the contributions of
`Weak Annihilation' (WA) to inclusive semileptonic decays of
heavy flavour
hadrons and the matrix elements of four fermion operators. We argue that
nonperturbative contributions from low energy hadronic final states provide
the dominant impact of WA on the semileptonic $b\ra u$ width
of $B^-$ mesons. In that case WA
affects the lepton spectrum mainly in its endpoint region --
i.e. beyond the kinematical boundary for $b\ra c$ transitions --
and we expect the
later to look quite different in $B^-$ than in $B_d$ decays.
At present we are unable to make a truly quantitative prediction; yet
a detailled experimental comparison of the $B^-$ and the $B_d$ lepton
spectra will enable one to determine the size of the
relevant matrix elements and to check directly to what extent the
factorization approximation works here.
Using universal analyticity properties of the
decay amplitude we analyze WA in the
perturbative regime and rederive our earlier results about
the absence of a power enhancement
due to gluon emission in a more general way.
We address also the problem of the
invariant mass of hadrons in the final state in semileptonic
$B$ decays and
identify nontrivial cancellations in $\aver{M^2\ind{hadr}}$.

\newpage
\large
\addtocounter{footnote}{-1}

\section{Introduction}

In previous papers \cite{we} we have
shown how the inclusive decays of heavy
flavour hadrons can be treated reliably in QCD via an expansion in
$1/m_Q$, $m_Q$ being the mass of the heavy flavour quark. The leading
term in this expansion represents
-- not surprisingly -- the results of
the simple spectator ansatz. We have found that there are
corrections of order $1/m_Q$ to the {\em shape} of spectra, but not
to totally integrated rates. Those receive corrections first on the
$1/m_Q^2$ level and they are of an intriguing nature:
they distinguish between baryon decays
on one hand and
meson decays on the other while affecting the latter in a
way that is practically independent of the flavour
of the spectator antiquark.
On the $1/m_Q^3$ level the
conventional spectator-dependent effects arise, namely
`Weak Annihilation' (WA),
`Pauli Interference' (PI) and `Weak Scattering' (WS) in
baryons\footnote{ Quite often a distinction is made
between the exchange of a
$W$ boson in the $s$ and in the $t$ channel referred to as Weak
Annihilation and $W$ exchange, respectively.
We find
such a classification somewhat artificial because both interactions lead
to very  similar four fermion couplings; they may differ in
their colour flow at {\em tree} level, yet this distinction fades away
under gluon renormalization.
For that reason we have adopted the
classification proposed in ref.~\cite{VS}:
WA denotes then the exchange of $W$ bosons
both in the $s$ and $t$ channel
of a quark-antiquark sytem whereas WS refers to $W$ exchange driving
quark-quark scattering in baryons.};
these represent preasymptotic corrections
to a parton model treatment whose size depends on the flavour of the
`spectator' quarks or antiquarks in a given heavy flavour hadron $H_Q$.

It is the purpose of this note to study in more detail
the role played by such spectator dependent effects; we will analyze
almost exclusively WA with the emphasis on
nonperturbative effects that may eliminate
chirality suppression and thus
enhance its effect to an appreciable level.
The discussion will center mainly on
semileptonic decays since those allow
a more reliable and more detailled theoretical
treatment. Among other things
one can go beyond total rates and study lepton
{\em spectra}. An analysis of
their observed shape will, at least in principle,
allow to extract the size of
hadronic matrix elements that in turn determine the
impact of WA on other quantities.

The paper will be organized as follows: after
a general overview of the effects associated with WA
we set the framework for
our theoretical discussion in Section 2 where we establish
the connection between the effects of WA and matrix elements
of four fermion operators that appear in
quark diagrams on the tree level. Such a relationship was
implicitly assumed in earlier
papers~\cite{VS,VSlog}; its existence was
later explicitely conjectured
and supported by the perturbative analysis of
ref.~\cite{BU}. Here we present a general analysis focussing on
nonperturbative effects. It was shown in ref.~\cite{BU} that the power
enhancements produced by gluons cancel each other in {\em inclusive}
rates. We illustrate the general formalism
adopted here by using it to
rederive these results; we demonstrate that they are indeed
-- as stated in ref.~\cite{BU} -- of rather general nature. In
Section 3 we apply these findings to the
semileptonic spectra and show that
the interesting effects originate in the endpoint region and
occur mainly there.
Still there are `tail' contributions coming from the endpoint effects;
in fact they
reflect the `hybrid' renormalization of the operators that was first
discussed
for inclusive effects in ref.~\cite{VSlog}.
In Section 4 we treat the invariant mass distribution of the hadronic
final state before summarizing our study in Section 5.
We will discuss primarily beauty decays with some comments added
on charm decays. It is worth noting that the first, though rather general,
study of the analyticity properties of the amplitudes for
semileptonic heavy flavour decays
has been done in ref.~\cite{CGG}; some details omitted in the
present paper can be found there.

WA must be understood in a more general way \cite{BU}
than the one
suggested by simple parton level diagrams. A difference in the lifetimes
of $B_d$ and $B^-$ mesons can be
produced by PI as well as by WA.
PI requires the presence of at least two identical quarks or
antiquarks. WA on the other hand is
associated with the presence of a
quark-antiquark pair of the same flavour;
the antiquark is originally present in the $B$ meson as a constituent
and the quark gets
created at the weak interaction vertex. In
semileptonic B decays PI cannot intervene and the whole difference in
$\Gamma _{SL}(B^-)$ vs. $\Gamma _{SL}(B^0)$,
according to the existing classification,  has to come from WA via the
KM suppressed $b\ra u$ transitions.

The simplest annihilation process is shown in Fig.1a and is
easily calculated; we will refer to it as `parton annihilation'.
It scales
like $(\hscale/m_b)^3$ with the mass of the heavy quark.
Yet rather subtle dynamical effects have to be addressed once one goes
beyond the simplest diagram.
For example $\bar qq$ pairs can annihilate into a
gluon, Fig.1b; the emergence of such new final
states in the decay represents a conventional annihilation effect.
However this is not
the only modification -- for annihilation affects at the same time
the production rates for
the `old' states as illustrated in Figs.1c: the original `quasifree'
amplitude now interferes with the annihilation-induced amplitude. This
modification does not represent the square of any particular amplitude and
therefore can actually be negative thus  {\em decreasing} the width.
Nevertheless
this latter kind of effect has to be included
under the term  WA unless one
wants to coin a new name for such spectator-sensitive corrections!

The amplitudes for a final state with a gluon or with
a $\bar qq$ pair when taken {\em separately} are enhanced by
inverse powers of light quark masses
(or more generally energies $\aver{E_q}$)
as compared to the parton estimate of WA.
This enhancement~\cite{BSS,VSlog,BU} can be
understood as being caused by final state interactions, namely the
mixing between nearly mass degenerate flavour
singlet $\bar qq$ pairs and
gluons\footnote{N.U. is grateful to J.Collins for the
discussion of this point.}. It is quite natural to expect
that this enhancement evaporates when one sums over these
decay processes. It was shown in ref.~\cite {BU} by explicit
calculation that this is indeed the case. It means by the way that the
two possible understandings of WA are dramatically different when the
decaying quark is heavy enough.

Of course the physical states coming from flavour singlet
$\bar qq$ pairs in practice cannot be distinguished from those
generated purely by gluons; i.e. observable
transitions cannot exhibit an $1/\aver{E_q}$ enhancement. The
decays of the $B_c$ meson consisting of $b\bar c$
offer an intriguing exception to
this general rule: for final states containing a $c$ and $\bar c$
pair
can be distinguished from those without heavy quarks
and the conventional
WA reaction $B_c\ra d\bar ug$ can thus be isolated.
On the other hand the overall impact of annihilation in our
more general formulation is hard to extract: for one cannot
any longer rely on a comparison of $\tau (B_c)$ vs.
$\tau (B_u)$ as an unambiguous yardstick since due to
$m_c \gg m_u$ no (light) flavour symmetry can be invoked.
There is however a saving grace:
both constituents of the $B_c$ meson are heavy and
therefore we know how to
explicitely compute the relevant transition rates.
The situation in $B_s$ decays
falls somewhere in between the decays of $B_c$ and $B_{u,d}$ mesons.
$SU(3)$
breaking effects in total beauty decay widths are under
control here~\cite{BUV,BUR}; yet at the same time
a reliable isolation of $s\bar{s}$ states
produced (or affected) by WA is questionable if
possible at all in a real experiment.

The analysis of ref.~\cite {BU} actually yielded a threefold
result concerning WA:

$\bullet$ All gluonic enhancements
of the form $(1/\aver{E_q})^n$ that appear in individual diagrams
cancel out in the inclusive transition rates.

$\bullet$ The emission of
gluons carrying momenta larger than the typical
hadronic scale $\mu\ind{hadr}$ provides the  `hybrid'
renormalization~\cite{VSlog} of the local four fermion operators
that drive the spectator-dependent effects in
the decays.

$\bullet$ The chirality suppression still persists
for WA when treating gluon radiation perturbatively
in the leading {\em log} approximation.

The last observation implies that the weight of
WA is reduced in heavy flavour meson decays,
but not necessarily to insignificant levels. For helicity
suppression could be overcome or its impact at least be
softened by ${\cal O}(\al_s)$ subleading
contributions from the emission of hard gluons, or via
nonperturbative effects. Numerically significant effects can be
expected from the latter, and we will discuss them now in detail
for the case of semileptonic $b\ra u l\nu$ transitions.

\section{General Framework; Total Rates}

This Section consists of four main parts:
(A) We will state
a relationship between the contributions of WA to
inclusive rates and
the matrix elements of four fermion operators
as it arises
in the approach formulated in
refs.~\cite{VS,VSlog}.
(B) We will discuss the contributions to WA from low mass
hadronic final states.
(C)  We will analyze how higher energy states contribute to WA
and show that their impact is given by the corresponding perturbative
corrections to the matrix elements of these operators.
(D) Some consequences for nonleptonic decays
will be pointed out.

{\bf (A)} Following the notation of ref.~\cite{CGG} where
the traditional approach to deep
inelastic scattering was modified for the case of semileptonic heavy
flavour decays,
one starts from the tensor
$$R_{\mu\nu}(q)=\int d^4z \exp(-iqz) \cdot
\matel{B}{iT(J_\mu^+(z)J_\nu(0))}{B}\:=\:-\delta_{\mu\nu}R_1+$$
$$+p_\mu p_\nu R_2
+i\epsilon_{\mu\nu\rho\lambda}p_\rho q_\lambda R_3 +
q_{\mu}q_{\nu} R_4 +
(p_{\mu}q_{\nu} +q_{\mu}p_{\nu}) R_5 +
i(p_{\mu}q_{\nu} -q_{\mu}p_{\nu}) R_6 \;\;;\eqno(1)$$
$q$ and $p$ are the momenta carried
by the
lepton pair $l\,\nu$ and the $B$ meson,
respectively; $J_{\mu}$
denotes the underlying quark current. (One has
$J_{\mu}=\bar b\gamma _{\mu}(1-\gamma _5)u$ when studying WA in
semileptonic $B$ decays.)
The scalar quantities $R_i$ are functions of the Lorentz
invariants $q^2$ and $x=1-(p\cdot q)/M_B^2$.
In the $B$ restframe $(1-x)M_B$ is
the energy of
the lepton pair whereas $x M_B$ represents
the energy of the final hadronic state\footnote{For convenience we have
adopted a different definition for the energy variable than used in
ref.~\cite{CGG}.}.  To compute decay rates one contracts the absorptive part
of $R_{\mu\nu}$ with the corresponding leptonic tensor $L_{\mu\nu}$ built
from the leptonic charged currents. The absorptive part of $R_{\mu\nu}$,
denoted by $W_{\mu\nu}$, is given by the discontinuities of the functions
$R_i$ in the complex variable $x$ in the physical region. For example the
total semileptonic width is given by
$$\Gamma\ind{sl} \propto \int_{0} dq^2
\int_{\frac{1-q^2/M_B^2}{2}}^ {1-\sqrt{q^2/M_B^2}} dx
\sqrt{(1-x)^2M_B^2-q^2}
\cdot \tilde{W}(x,q^2)\;\;,$$
$$\tilde{W}(x,q^2)=q^2 W_1(x,q^2)
+\frac{M_B^2}{3}(M_B^2(1-x)^2-q^2)W_2(x,q^2)=
\frac{1}{3}(q_\mu q_\nu-q^2\de_{\mu\nu})W_{\mu\nu}(q)\;\;
\eqno(2)$$
where Eq.(2) has been written for the case of massless
leptons\footnote{Note the explicit nonanalytical factor
$\sqrt{(1-x)^2M_B^2-q^2} = |\vec{q}|$ that did not appear
in the discussion of ref.~\cite{CGG}.}.

To establish a relationship between the correlator
function of eq.(1) and the
local product of the corresponding quark currents, we start
from a general expression for the time-ordered
product of any two operators:
$$\int d^4z \exp(-iqz) \cdot \matel{B}{iT(O_1(z)O_2(0))}{B}\:=$$
$$=\:\sum_n
\frac{\matel{B}{O_1(z_0=0)}{n}\cdot
\matel{n}{O_2(0)}{B}}{-E_B+E_n+q_0-i\epsilon} +
\sum_{n'}\frac{\matel{B}{O_2(0)}{n'}\cdot
\matel{n'}{O_1(z_0=0)}{B}}{-E_B+E_{n'}-
q_0-i\epsilon} \;\;,\eqno(3)$$
where $n$, $n'$ are intermediate states with energies $E_{n,n'}$ and
$E_B$ denotes
the energy of the $B$ meson. Integrating the imaginary part of eq.(3)
over the
energy $q_0$ from $E_B-E_2$ to $E_B-E_1$ yields
$$\int_{E_B-E_2}^{E_B-E_1} dE \frac{1}{\pi}
\Im\matel{B}{iT(O_1O_2)}{B}_{E,
\vec{q}}\;=$$
$$=\;\sum_{E_1<E_n<E_2}\;(2\pi)^3
\de^3(\vec{p}_n-\vec{p}_B+\vec{q})\;
\matel{B}{O_1(z_0=0)}{n}\cdot\matel{n}{O_2(0)}{B} + $$
$$+ \sum_{2E_B-E_2 < E_{n'} < 2E_B-E_1}\; (2\pi)^3\de^3(\vec{p}_{n'}-
\vec{p}_B-\vec{q})\;\matel{B}{O_2(0)}{n}
\cdot\matel{n}{O_1(z_0=0)}{B}\; . \eqno(4)$$
Since
$$\matel {B}{J_{\mu}^+(z_0=0)J_{\nu}(z_0=0)}{B}_{\vec{q}}= $$
$$ = \;(2\pi)^3\sum_{n}
\de^3(\vec{p}_n-\vec{p}_B+\vec{q})\;
\matel{B}{J_{\mu}^+(z_0=0)}{n}\cdot\matel{n}{J_{\nu}(z_0=0)}{B}\;\;
\eqno(5)$$
one arrives at the following conclusion:
integrating
the discontinuity of the Greens function in eq.(1)
over some finite energy interval yields the contribution to
the matrix element of
the {\em equal time} product of the hadron currents from
intermediate states having kinematically allowed
energies for this energy interval.
This relationship holds as long as $u$ channel
singularities (at $x\, >\, 1$) in the time ordered Greens functions do not
contribute in the energy interval considered.
Integrating also over the space
momenta $\ve{q}$ one finally obtains a completely
local product of currents.

After this general remark we return to the semileptonic width
as given by eq.(2). The various
intermediate states saturating the correlators $W$ are
characterized by rather
different energy and momentum scales;
the dynamics associated with them is therefore quite different
and we will
discuss them separately. For the sake of definiteness
we shall assume in what
follows that we are in the rest frame of the $B$ meson (or one where its
motion is at most nonrelativistic). We anticipate the major
contribution to WA to come from
nonperturbative effects, and to show up in low
lying states. So we shall start with discussing the range of low $E_n\sim
\hscale$, i.e. $x\sim \hscale/m_b \ll 1$.

{\bf (B)} When both the energy and the momentum of the hadronic
final state are small compared to $M_B$ the lepton
pair in the decay carries a large invariant mass, $q^2\simeq m_b^2$,
making the leptonic part of the amplitude (including the $l\,\nu$ phase
space factor) basically insensitive to both the energy and the momentum
of the state $\state{n}$. Its value can thus be
evaluated at
$q_\mu \simeq p_{\mu}=(M_B, 0,0,0)$; summing over low lying states
with a constant weight factor given by the leptonic tensor
$L_{\mu \nu}(p)$ one
actually performs an integration of the functions $\tilde{W}$
over both energy and
momentum . The hadronic part of the decay probability is thus
indeed determined by the matrix
element of the local operator
$\bar{u}_L\gm b_L(0)\cdot \bar{b}_L\gn u_L(0)\;$.
We then draw the following conclusion:
{\em The width for semileptonic decays into hadronic final states
with energies
substantially
smaller than $M_B$ is described by the contributions of
these states to the matrix element of the local product of the
underlying weak quark currents, taken between $B$ meson states}~--
$\matel{B}{\,\bar{u}_L\gm b_L(0)\cdot \bar{b}_L\gn u_L(0)\,}{B}$ --
{\em and
multiplied by the leptonic tensor evaluated at} $q^2=M_B^2$.
Corrections due to the small (for $E_n\ll m_b$ )
variation in the lepton pair momentum
can be taken into account by expanding the leptonic
part of the amplitude around $q^2=M_B^2$. They are
described by similar four fermion operators containing
derivatives; their impact is suppressed by extra powers of $\hscale/m_b$.

Since our focus is on width differences due to WA we
consider the difference of the matrix elements taken between
charged and neutral mesons:
$$M_{\mu\nu}=4(\matel{B^-}{\bar{u}_L\gm b_L\cdot \bar{b}_L\gn u_L}{B^-}-
\matel{\bar{B^0}}{\bar{u}_L\gm b_L\cdot \bar{b}_L\gn u_L}{\bar{B^0}})=
f_B^2(v\,p_{B_\mu}p_{B_\nu} - g\,M_B^2 \delta_{\mu\nu})\;\; \eqno(6)$$
i.e. the size of WA is expressed in terms of two dimensionless
parameters $v$ and $g$.

The contributions of WA to the width
are obtained, as already stated, by
contracting $M_{\mu\nu}$ with the
leptonic tensor $L_{\mu\nu}(p_B)$
given by the absorptive part of the $l\nu$ loop
$$L_{\mu\nu}(q)= - \frac{1}{6\pi} (1-m_l^2/q^2)
[\;(1-\frac{m_l^2}{2q^2}-\frac{m_l^4}{2q^4})\;q^2\de_{\mu\nu}\; -\;
(1+\frac{m_l^2}{q^2}-
\frac{2m_l^4}{q^4})\;q_{\mu}q_\nu\;]\;\;\eqno(7)$$
where $m_l$ denotes the charged lepton mass; the correlator of
the leptonic charged currents $L_{\mu\nu}$ of course has a
transverse structure for $m_l=0$. For the
contribution of WA to semileptonic
B decays one then finds
$$\Delta\Gamma\ind{SL}=\Gamma _{SL}(B^-)-\Gamma _{SL}(B^0) \simeq
\frac{G_F^2|V_{ub}|^2 f_B^2 M_B^3}{8\pi}
(1-\frac{m_l^2}{M_B^2})(v\frac{m_l^2}{M_B^2} + 2g) \eqno(8)$$
The part of the hadronic matrix
element which is proportional to $p_{\mu}p_{\nu}$
yields a result that vanishes in the limit of
$m_l\rightarrow 0$: this reflects the
conventional chirality suppression
of WA as it applies to semileptonic B decays.
On the other hand the part in $M_{\mu\nu}$ proportional to
$\de_{\mu\nu}$ does not suffer from such a reduction; its weight
then depends on the size of $g$. In general $g\neq 0$  and therefore
the impact of WA on semileptonic decays $b\ra u$ could be
significant as seen
from the following comparison:
$$\frac{\Gamma _{SL}(B^-)-\Gamma _{SL}(B^0)}
{\Gamma (B\ra \tau\nu_\tau)}\simeq
\frac{2gM_B^2}{m_\tau^2}\; \sim \; 6
\; \; \; \mbox{for} \; \; \; g\; \sim \; \frac {1}{3}
\; \; \; .\eqno(9)$$

The simplest annihilation decay is $B\ra l\nu_l$; it
is represented by $\state{n}=\state{0}$ in the current correlator
decomposition of eq.(3) and correspondingly
by the vacuum factorization ansatz for the matrix
element $M_{\mu\nu}$, eq.(6), of the local four fermion operator.
The contribution of
the decay $B\ra l\nu_l$ to $\Gamma (B^-)-\Gamma (B^0)$ is
therefore given exactly by the
factorizable piece of the
underlying matrix element, i.e. in the notation of eq.(6) by
$$v\ind{vac}=1\;\;\;,\;\;\;\;\;
g\ind{vac}=0 \;\;.\eqno(10)$$
Helicity suppression applies and the dominant mode in this
category is $B\ra \tau \nu$.
Yet in general $g\neq 0$ will hold for the matrix element in eq.(6)
and helicity suppression
can thus be expected to be vitiated
when hadrons are present in the final state; the relevant question
is on which {\em numerical} level will this arise.

{\bf (C)} Up to this point we have discussed low energy
hadronic final states where
a perturbative treatment is never applicable;
their combined effect is to
generate a (nonperturbative) value for the quantities $v$ and $g$.
When higher momenta arise  in the
hadronic final state one can consider analyzing
the structure of inclusive
hadronic states and estimate their contributions by using
perturbative QCD
that operates in terms of quark and gluon fields.
Such an  analysis was first undertaken in a similar context in
ref.~\cite{BSS} with the focus mainly on overcoming
helicity suppression. Only the one-gluon intermediate
state was considered there: its amplitude was found to be free of
helicity suppression and in addition to be enhanced by a factor of
$m_b^2/\aver{E_{sp}}^2$ relative to the WA width without gluons,
with $\aver{E_{sp}}$ being an average energy of the light
(anti)quark. Yet in
ref.~\cite{BU} we have shown that such an enhancement disappears from
the {\em fully inclusive} amplitude, i.e. the latter is {\em regular}
in the limit of vanishing spectator quark momenta.
We actually found that for $\hscale  \ll  E \ll m_b$
(with $E$ denoting the energy of the hadronic states)
the WA contribution is of the form
$\al_sdE/E$.
Upon integration over this energy range it contributes to
the so-called `hybrid' renormalization of the
underlying four fermion operator. In other words,
summing over all hadronic final states leads to the renormalization of
the operator entering eq.(6) down to the
low scale $\sim \hscale^2$.

An explicit calculation has shown -- somewhat surprisingly --
that the hybrid renormalization does {\em not}
modify the Lorentz structure of the four quark operator.
Therefore perturbative corrections {\em per se}
do not induce a $g$ term for
the matrix element in eq.(6) in the
leading log aproximation (LLA)
and helicity suppression is thus found to persist on that level!
It can be overcome only
at a relatively small level ${\cal O}(\al_s(m_b^2)/\pi)$ beyond LLA
-- unless
it emerges via nonperturbative dynamics as was discussed above.

It is instructive to see schematically how the cancellation of the
$1/\aver{E_{sp}}^2$ and $1/\aver{E_{sp}}$
terms occurs and the stated structure of the
perturbative corrections emerges by using rather general
considerations which are based on
the analytical properties of the correlator in eq.(1).
It will demonstrate -- as was stated in ref.~\cite{BU} -- that
the underlying argument is more general than the simple quark diagram
that is encountered in the explicit calculation of the
perturbative corrections;
it will be of use for our subsequent discussion of semileptonic
spectra.
For the sake of simplicity we will ignore here the colour quantum
numbers that strictly speaking forbid
these gluonic effects to first order in $\al_s$ for semileptonic decays;
in fact the diagrams we consider are literally those
that were treated in ref.~\cite{BU} in addressing
nonleptonic decays.

The first nontrivial perturbative correction to the annihilation width
is obtained for example from the diagram in Fig.2;
we consider it in greater
detail because it could {\em a priori}
lead to the most singular corrections.
The width given by this diagram is most
conveniently obtained by keeping the value of $q^2$ at first fixed and
integrate
over $x$, i.e. over the energy; subsequently one performs
the integration over $q^2$ according to eq.(2). This diagram generates
the correlator function $R$ with the following obvious
analytical properties in
the complex $x$ plane:
there are three
poles corresponding to the massless gluon and to on shell $\bar u u$
states with $\bar u$ being the spectator
and $u$ coming from the $b\ra u$
transition. These $\bar uu$ pairs
have an invariant mass squared of
$(k+p\ind{sp})^2\simeq 2(k \cdot p\ind{sp})$.
This is much smaller than $m_b^2$ even for a high
momentum $|\ve{k}| \sim m_b$.
The three poles are therefore very closely spaced in the
$x$ plane:
$$x_1 \simeq \frac{1-Q^2}{2}\;\;,\;\; x_{2,3}\simeq
\frac{1-Q^2}{2}+\frac{(p_u \cdot
p\ind{sp})}{m_b^2} \;\;\eqno(11)$$
as indicated in Fig.3; we have introduced here the dimensionless
ratio $Q^2= q^2/m_b^2$.
Integrating the imaginary part of this diagram
(which in this particular case amounts to summing the
three $\de$-functions) can be done by evaluating
the integral over path $\al$ in Fig.3.
The standard procedure
is then to move the integration contour into the complex plane
away from the
singularities, like path $\be$. As long as that countour does not
approach the singular points $x$ near $(1-Q^2)/2$
one can safely set the momenta of the spectator quark
to zero -- for it merely shifts the exact position of the
poles by a small
amount. This order $\alpha _S$ contribution to the
total width therefore possesses a regular limit for vanishing spectator
momenta;
it can be calculated by integrating the function
containing the triple pole at the point
$x\simeq (1-Q^2)/2$ over a contour like $\beta$.
Accordingly the decay amplitude cannot contain any inverse powers of the
spectator mass or energy~\cite{BU}. The explicit
evaluation of this diagram is thus in principle
straightforward; yet in practice it requires attention
to technicalities like the numerators in the fermion
propagators that restore the correct dimensionality for this
diagram.
They have the generic structure $(p_b-q)_\al(p_b-q)_\be$;
therefore they either cancel one of the
three poles $1/(1-Q^2-2x)$ in the amplitude or instead
provide a factor
$\vec{q}\,^2=((1-x)^2-Q^2)m_b^2$ that will vanish at the point
$\sqrt{Q^2}=1-x$ where the Jacobian in eq.(2) is singular.
An explicit expression
is easily obtained by stretching the integration contour
around the cut produced by the Jacobian $\sqrt{(1-x)^2-Q^2}$ in eq.(2) as
represented by the contour $\ga$ in Fig.3; alternatively one can use
the relation
$$\int dx f(x) \frac{1}{\pi} \Im \frac{1}{(x-x_o+i\epsilon)^{n+1}} =
-\frac{1}{n!} f^{(n)}(x_o)\;\;.$$
Upon differentiating the Jacobian $\sqrt{(1-x)^2-Q^2}$ one obtains
for both expressions of the numerators in the propagators a
result $\propto \frac{1}{1-Q^2} = M_B/2|\vec{k}|$ where
$\vec{k}$ is seen as the momentum of the massless
quark in the semileptonic decay at a given value of
the lepton invariant mass
$q^2$. The expression
indeed becomes singular when the
singularity of the function pinches the cut
of the Jacobian corresponding to the edge of phase
space, i.e. at the maximum invariant mass of the lepton pair.
The integration over $q^2$ in this region
produces the hybrid renormalization $log(m_b^2/\mu ^2)$.

So far we have discussed WA in terms of the annihilation of
free quarks. Yet one needs the
matrix element of the current correlator between $B$ hadrons, eq.(1).
Let us consider the
matrix element
of the one gluon annihilation transition between a real $B$ meson.
Due to the distribution in the
relative motion of the spectator (anti)quark and the
heavy quark inside the hadron the three original poles in the
amplitude get smeared out and actually turn into cuts.
Higher order corrections are expected to produce a similar effect.
Yet the general arguments proceed as before as
long as the invariant mass of the intermediate state is small compared to
$m_b$ (or more exactly to the energy released into the final hadronic
states). For
this ensures that the singularities are shifted from their original
position $x_o=(1-Q^2)/2$ by a much smaller amount than
its distance to the Jacobian singularity
$x=1-\sqrt{Q^2}$ which reflects the edge of phase space.
The smallness of the mass of the intermediate state is the only essential
fact one has to invoke
from a quark model description of the heavy hadron; in the naive constituent
quark model this mass squared indeed cannot
exceed the scale $\hscale\cdot m_b$.
As long as such an inequality holds and the momenta of the final hadronic
states are much smaller than $m_b$
one can deduce as before that the resulting
contributions to the decay width of the
hadron are given by the corresponding corrections
to the matrix element of the
local product of currents. This is the reason why the perturbative
effects of the one gluon annihilation process are
reproduced correctly by
calculating the corresponding
{\em Euclidean} graph for the {\em local} four fermion operator~\cite{BU}.

The situation would appear to change somewhat when higher
order corrections are
considered; in particular the pole corresponding to the gluon propagator
changes drastically and the integral over the invariant mass of the
gluon fragmentation will extend over a wider range .
None of this will however spoil the cancellation of the enhanced terms
$\propto m_b^2/\aver{E_{sp}}^2$ and $m_b/\aver{E_{sp}}$.
For the amplitude
computed via the contour $\be$ in Fig.3 is determined
by momenta that are high compared
to $\hscale$; the overall effect is therefore determined by the ultraviolet
properties of the parton diagram rather than by details of the real hadronic
states (as long as the invariant mass of the lepton pair does not approach
the mass of $b$ quark too closely).

{\bf (D)} The previous discussion provides the formal framework
for the QCD treatment of
WA in semileptonic decays.
A similar analysis -- although of course not so
rigorous -- can be applied to nonleptonic
heavy flavour decays as well with obvious technical modifications: one starts
from the correlator function generated by the four fermion Lagrangians rather
than the weak currents in eq.(1). To obtain
total widths one
considers the correlator at $q=0$
and applies an operator expansion in $1/m_b$ to it.
The leading spectator dependent
effects are obtained when one treats a quark-antiquark (for WA) or
a quark-quark (for PI or WS) pair in the intermediate state
as free fields. The
general structure thus obtained resembles very much that found
in semileptonic decays; in particular the analogous four fermion
operators appear then, albeit with a somewhat more general
Lorentz and colour structure. Subleading corrections
-- both perturbative and nonperturbative ones -- to nonleptonic
and semileptonic rates do differ and {\em a priori} are expected
to be larger in
the former than in the latter. Yet it is important that they reflect
the same underlying structure;
their relative weight both for perturbative and nonperturbative
corrections is controlled by the same quantity, namely the distance
from the thresholds of the
free parton model graph.

Our preceeding discussion has yielded two important phenomenological
conclusions:
{\bf (i)} The actual size of WA for semileptonic $b\ra u$ decays is indeed
determined by the matrix
element $\matel{B^-}{\bar{u}_L\gm b\cdot \bar{b}\gn u}{B^-}$;
nonleptonic decays are governed by similar matrix elements with
different colour contracting scheme\footnote{This
suggestion was formulated in ref.~\cite{BU} although it had been implicitely
implied already in analysis of ref.~\cite{VSlog}.}.
{\bf (ii)} Helicity suppression persists in the presence of {\em
perturbative} gluon corrections, even when `hybrid' renormalization is
included in the leading log approximation.

On the other hand helicity suppression can be expected to be vitiated by
subleading perturbative or by nonperturbative corrections.
This is to say in the notation of eq.(6) that in general a term
proportional to $\delta _{\mu\nu}$ will be induced, albeit with a
$g$ presumably somewhat smaller than unity.
Numerically sizeable WA contributions
to inclusive rates can presumably be expected
only from nonperturbative dynamics.
WA could then still
change the width of semileptonic $b\ra u$ decays in
$B^-$ by up to twenty percent relative to that in
$B^0$. Such an effect would not be of purely academic interest
due to the following observation: one would expect -- if
nonperturbative effects are the main driving force overcoming helicity
suppression -- that WA would populate semileptonic final states
with mainly low energy hadronic systems. WA would thus contribute the
bulk of its weight in the
relatively narrow endpoint region of semileptonic decays which in turn
would magnify its impact there.
In the next section we describe the impact of WA on semileptonic
{\em spectra} in more details.

\section {Lepton spectra}

The energy spectrum for the charged leptons in semileptonic
beauty decays can be expressed in terms of the correlation function
given in eq.(1):

$$\frac{d\Gamma}{dy}\;=\;\frac{G_F^2 |V_{ub}|^2}{32\pi^4}M_B^5\cdot
\int_0^{y} dQ^2\int^{\frac{1-Q^2}{2}+
\frac{1}{2}(1-y)(1-Q^2/y)}_{\frac{1-Q^2}{2}} dx
\tilde{V}(x,Q^2)\;\;=$$
$$=\;\;\frac{G_F^2 |V_{ub}|^2}{32\pi^4}M_B^5\cdot
\int^{1-y/2}_{(1-y)/2} dx\int_0^{2y(1-x-y/2)} dQ^2
\tilde{V}(x,Q^2)\;\;,$$
$$\tilde{V}(x,Q^2)=\frac{q^2}{M_B^2} W_1(x,q^2)
+(y(1-x-\frac{y}{2})M_B^2-\frac{1}{2} q^2)W_2(x,q^2) +
(1-x-y)q^2W_3(x,q^2)\;\; \eqno(12)$$
with $y=2E_{l}/M_B\:|\ind{c.m. frame}$ denoting the normalized charged lepton
energy; lepton masses have been ignored here.  The lepton spectrum in charm
decays is described by the same formula with two obvious modifications:
$M_B$ is replaced by $M_D$, and the positive sign in front of the $W_3$ term
in eq.(12) is changed into a negative one reflecting the $V+A$ structure of
the charge conjugated lepton current.  Integrating eq.(12) over $y$
reproduces of course eq.(2) for the total semileptonic width.

Different kinematical regions of this energy
spectrum are shaped by different
dynamical regimes controling the recoiling hadronic system.
As before we will first consider low energy states
affected by nonperturbative dynamics.
Low mass hadronic states whose energy does not exceed
some small value $M\ind{np}$ can contribute only to a rather limited slice
in the lepton energy: $1-2M\ind{np}/M_B \lsim y < 1$; this interval becomes
more and more narrow as the mass of the heavy quark increases
and actually shrinks to a
point in the limit of an infinitely heavy quark.
A simple minded valence quark description suggests that
$M\ind{np}\sim m_{sp}\sim \hscale$, i.e. this interval of the lepton
energy remains constant in absolute units.
On the other hand the main effect of WA as analyzed in the previous
section is to occur in this
small region! The contribution from WA then appears in the lepton energy
distribution effectively
like a $\de$-function near $y=1$ with a smearing in the energy of order
$\hscale/m_b$. (It is worth noting that similar effects have
been identified
in ref.~\cite{BSUV} for flavour-independent preasymptotic
corrections.) This contribution to the total width is thus concentrated in
the region A in Fig.4 and its overall strength is determined
by the value of $g$ in the matrix element eq.(6).
The {\em shape} of the spectrum
inside this region on the other hand depends on the details of
hadronization into the low-lying hadronic states.
At present neither of these can be calculated reliably from QCD.

If the energy released into the hadronic system
is large compared to the typical invariant mass of the latter
then one can rely on a
perturbative treatment. The quark model suggests for the
square of this mass as a typical value $M\ind{annih}^2\simeq
|\ve{q}|\cdot m\ind{sp}$; the aforementioned requirement is thus
indeed satisfied even for large three-momenta.
Accordingly we may rely on a perturbative treatment of
WA below region A.
(There are actually two subregions shown in Fig.4 -- B and C -- which
are separated by the point $y=1-M_D^2/M_B^2$; $b\ra c$ transitions
are restricted to region C.)
The structure of the perturbative amplitude can again be simplified by
considering it in the limit of vanishing spectator momenta; the
functional dependence of the result can then be readily obtained from
the discussion of the previous section. For example the contribution to
$\tilde{V}$ coming from the part
of the amplitude containing a double pole is given by
$\de'(\frac{1-Q^2}{2}-x)$, and
the integration over $x$ in eq.(12) yields merely
$\de((1-y)(1-Q^2/y))$; the integration over $Q^2$ then leads to

$$\frac{d\Gamma}{dy} \simeq c\frac{\al_s}{1-y}\;\;\mbox{ for }
\;\; 1-y > M\ind{annih}^2/m_b^2 \sim \hscale /m_b
\;\;, \eqno(13)$$
where the factor $c$ comes from the
``$\de$-function'' term in the spectrum region A.
Eq.(13) gives the hybrid logarithm
$\log{m_b^2/\hscale^2}$ when integrated over the whole spectrum and
in fact reflects the nonfactorizable contribution to the hybrid
renormalization of the operator in eq.(6). Factorizable
corrections describing (virtual) corrections to individual semileptonic
vertices also
produce such terms. Their effects however do not depend on the lepton energy
$y$; they lead to the known renormalization of both weak currents in
the matrix element eq.(6) down to the low scale $\hscale^2$. These in
principle increase the height of the spectrum in region A by a factor
$\appa^{4}$ (given explicitely below in eq.(14)) over the
naive quark model result. This has been incorporated
in eq.(6) via the explicit factor $f_B^2$ undergoing
the same corrections~\cite{VSlog}.
The WA contribution over the whole spectrum is sensitive to
both types of hybrid renormalization.

After this general overview of the structure
of the perturbative corrections we give explicit
expressions.
Four fermion operators as they appear in eq.(6)
do not change
their Lorentz structure upon hybrid renormalization~\cite{VSlog,BU}
as long as they are constructed from purely left-handed fields;
yet they mix
with the colour octet counterparts:
$$\bar{u}_L\gm b_L\cdot \bar{b}_L\gn u_L \;\;\ra\;\;
(\appa^{9/2}-1/9(\appa^{9/2}-1))
\cdot (\bar{u}_L\gm b_L) (\bar{b}_L\gn u_L)- $$
$$ - 2/3(\appa^{9/2}-1)\cdot (\bar{u}_L\frac{\lambda^a}{2}
\gm b_L) (\bar{b}_L\frac{\lambda^a}{2}\gn u_L)\;\;,$$
$$\appa=[\frac{\al_s(\mu^2)}{\al_s(p^2)}]^{\frac{1}{b}}\;\;,
b=11-\frac{2}{3} n_f\;\;. \eqno(14)$$
The impact of WA on the continuous part of the spectrum is thus
determined by the matrix elements of both colour singlet and colour
octet operators:
$$\int _y^{1} \frac {d\Gamma\ind{annih} (\eta)}
{d\eta } d\eta \simeq N \cdot
\appa^{-4}f_B^2\cdot(\frac{\appa}{\appa_y})^{4}
[ g\ind{singl}
(\appa_y^{9/2}-1/9(\appa_y^{9/2}-1))
-2/3 g\ind{oct} (\appa_y^{9/2}-1)]\,\,,$$
$$ N=\frac{G_F^2 |V_{ub}|^2 M_B^3}{4\pi}\;\;\;,\;\;\;
\appa_y=[\frac{\al_s(\hscale^2)}{\al_s(\, ([(1-y)m_b]^2\, )}]
^{\frac{1}{b}}\;\;\;,
\;\;\;\appa=
[\frac{\al_s(\mu\ind{hadr}^2)}{\al_s(m_b^2)}]
^{\frac{1}{b}}\;\; \eqno(15)$$
where we have used the notation
$$\matel{B^-}{\bar{u}_L\gm b_L\cdot \bar{b}_L\gn u_L}{B^-}-
\matel{\bar{B^0}}{\bar{u}_L\gm b_L\cdot \bar{b}_L\gn u_L}{\bar{B^0}}= $$
$$=\appa^{-4}f_B^2\cdot(v\ind{singl}\,p_{B_\mu}p_{B_\nu}
- g\ind{singl}\,M_B^2
\delta_{\mu\nu})\;\; $$
$$\matel{B^-}{\bar{u}_L\frac{\lambda^a}{2}\gm b_L\cdot
\bar{b}_L\frac{\lambda^a}{2}\gn u_L}{B^-}-
\matel{\bar{B^0}}{\bar{u}_L\frac{\lambda^a}{2}\gm b_L\cdot
\bar{b}_L\frac{\lambda^a}{2}\gn u_L}{\bar{B^0}} = $$
$$ =\appa^{-4} f_B^2 \cdot (v\ind{oct}\,p_{B_\mu}p_{B_\nu}
- g\ind{oct}\,M_B^2
\delta_{\mu\nu})\;\;.\eqno(16)$$
The quantities $\appa^{-4}$ have been factored out to make
the dimensionless
matrix elements $g$ and $v$ constant in the heavy quark limit.
The operators
in eq.(16) are normalizied at the low scale $\hscale$.
According to eq.(15) the total annihilation width and its part that lies
in the endpoint region A are given by, respectively
$$\Delta \Gamma\ind{annihil}\simeq \frac{G_F^2|V_{ub}|^2M_B^3}{4\pi}\cdot
(g\ind{singl}-0.4 g\ind{oct})\; \; \; ,$$
$$\Delta \Gamma\ind{annihil}|\ind{A} \simeq
\frac{G_F^2|V_{ub}|^2M_B^3}{4\pi}\cdot
g\ind{singl}\; \; \; ,$$
The numerical factor in front of $g\ind{oct}$ corresponds to the
normalization point $\al_S(\mu ^2)=1$ and $\Lambda\ind{QCD}=180\MeV$.

In Fig.4 we illustrate the spectrum of eq.(15) for the two typical
ratios $g\ind{singl}/g\ind{oct}=-1$ and $g\ind{singl}/g\ind{oct}=1/3$;
if the matrix elements have opposite signs the overall effect of
annihilation is actually enhanced relative to the one in the endpoint
refion A. Finally we have added in Fig.4 a $\de$-function at $y=1$
representing the two body decay $B^{\pm}\ra l \nu$; its finite height is
designated to reflect the chirality suppression of this mode.

they had the same relative sign the
overall effect of annihilation would
be smaller than
the one in the end point region A.
Finally we have added in Fig.4 a $\de$-function at
$y=1$ representing the
two body decay $B^\pm \ra l\nu$ (its 
finite height is designated to reflect the
chirality suppression of this mode) and
{\em ad hoc} some small nonleading perturbative corrections.

Turning to the discussion of real $B$ mesons we have to concede
that the numerical relation between
the effects in the perturbative (B,C) and
nonperturbative (A) regions depends on the
concrete position of the borderline
$\hscale$ between the two regimes. In other words
we do not know reliably how
close numerically we can approach the endpoint while
still observing the
perturbative $dy/(1-y)$ behaviour.

In reality the scale $m_c^2/m_b$ that determines the energy where
semileptonic $b\ra c$ decays start to contribute
is rather close to the
hadronic scale $\hscale$ (see e.g. the discussion in ref.~\cite{BSUV}).
Therefore only a small interval is left above the
charm threshold for the perturbative regime. This raises legitimate
doubts whether our treatment in its
present stage allows for a detailled
quantitative analysis of the regions A and B {\em separately}.

\section{Invariant Mass of the Hadronic Final State}

Understanding the structure of the hadronic final state is
obviously of great theoretical interest;
in addition it provides important
help in reducing the background in experimental studies\footnote{We are
grateful to Sheldon Stone for emphasizing this point to us.}.

The expectation value for the invariant hadronic mass (squared)
provides the simplest yardstick for the structure of the final state.
There are perturbative contributions to it from gluon bremsstrahlung;
they are infrared safe and thus calculable in a straightforward manner
(see e.g.~\cite{ACM}). The situation becomes considerably
more delicate for nonperturbative corrections. Although the
{\em shape} of distributions -- say the lepton energy spectra in
semileptonic decays -- generally gets $1/m_b$ corrections due to
the Fermi motion~\cite{BSUV} of the b quark, {\em integrated} rates do not
-- provided they are expressed in terms of the {\em quark} rather
than the {\em hadron} masses. To say it differently:
while the leading corrections are indeed of
order $1/m_b$ they can be expressed completely in terms of
$\bar\Lambda$ -- the asymptotic mass difference
between the hadron and the heavy quark (for details see
ref.~\cite{BSUV2}).

At this point we disagree with the conclusions of
ref.~\cite{CGG}: whereas allowance was made there for
$1/m_b$ nonperturbative corrections to the total
semileptonic widths, it was claimed that no such corrections arise in
either the shape of the lepton spectrum or in the average hadronic
invariant mass. The oversight in the latter claim is easily traced:
the physical invariant mass in the decay is given by
$$M\ind{hadr}^2=(p_B-q)^2=M_B^2 (2x-1+Q^2) \eqno(17)$$
where $x$ and $Q^2$ are as defined in Section 2, i.e. in terms of the
mass and the four momentum of the beauty {\em hadron}.
On the other hand,
as shown in ref.~\cite{we} no
$1/m_b$ terms are present in the integrated distributions
once they are expressed in terms of the mass and the four
momentum of the beauty {\em quark}. Nontrivial corrections
appear at the $1/m_b^2$ level; their general expressions are given
in ref.~\cite{BKSV}. Therefore it is the quantity
$$\aver{(p_b-q)^2}=\aver{M\ind{hadr}^2}-2\aver{(p_b-q)(p_B-p_b)}
-\aver{(p_B-p_b)^2} \eqno(18)$$
rather than the invariant mass that is not
subject to $\hscale m_b$ corrections;
on the other hand the leading $1/m_b$ corrections to the invariant mass
squared of the hadrons in the final state are given by
$$\langle M\ind{hadr}^2 \rangle\simeq m_q^2 + \frac{z_m(x)}{z_0(x)}
\bar{\Lambda} m_b\;\;,$$
$$ z_m(x)=
\frac{7}{10}-\frac{5}{2}x+16x^2-16x^3+\frac{5}{2}x^4-\frac{7}{10}x^5-
6x^2(1+x)\log\frac{1}{x}
\;\;\;,\;\;$$
$$z_0(x)=
1-8x+8x^3-x^4+12x^2\log\frac{1}{x}\;\;\;,
\;\; x=(m_q/m_b)^2 \;\;\;.\eqno(19)$$
The numerical factor $z_m/z_0$ varies
from $.7$ to $2$ for $m_q$ between $0$ and $m_b$; it gives just the average
energy of the quark $q$ in the free semileptonic decay, measured in terms of
$m_b/2$. This resolves the apparent puzzle raised in ref.~\cite{CGG}:
the absence of
operators yielding $1/m_Q$ corrections was interpreted there as implying
that no $1/m_Q$ contributions arise in the invariant
hadronic mass of the final
state -- although very intuitive physical arguments suggested
otherwise.
Note also that eq.(19)
satisfies the heavy quark limit when $m_b-m_c\ll m_c$: it yields
the mass $(m_c+\bar\Lambda)^2\simeq M_D^2$.

We will study here in particular how $\langle M_{hadr}^2\rangle$ differs in
$B_d$ and $B^-$ decays. The effects of perturbative
gluon bremsstrahlung in the
quasifree amplitude cancel out in such a difference, hence the
nonperturbative contributions are well defined here.

The contributions of WA to the total width scale like $1/m_b^3$;
it would then be natural to conjecture that the shift in $M\ind{hadr}^2$
caused by WA is rather
suppressed, namely of order $\hscale^3/m_b$ or even $\hscale^4/m_b^2\;$.
This would be in agreement with the observation that
$\aver{M_{hadr}^2}$ is not or hardly modified by
$B\rightarrow l\,\nu_l$ and other decays into final states
with low hadronic energies as described by the matrix
elements of eq.(6), because $M^2\ind{hadr}$ is very small in such
transitions. On the other hand gluon emission changes
the kinematics completely.
Strictly speaking colour conservation requires the presence of at least
two gluons for WA driven semileptonic decays.
For the sake of
simplicity we again ignore the colour indices and will discuss
the general effect for the case of one gluon in the intermediate state.

Two observations would seem to lead to a startling conclusion: the
processes with a real gluon, Fig.1b, and the one
containing an on shell $u\bar u$
pair produced via a virtual gluon, Figs.1c -- when considered separately --
induce large
corrections to the decay probability $\propto 1/m_b\;$, see e.g.
ref.~\cite{BU}.  Whereas the hadronic mass for the real gluon annihilation
of Fig.1b is literally zero, the typical invariant mass square for two other
cuts is $m\ind{sp}\cdot m_b$ because the main contribution comes from
$\ve{k}_u\sim m_b$. Yet the latter produce a (negative) correction to the
width of order $1/m_b$! This would imply that through gluon emission WA can
induce a negative correction to $M\ind{hadr}^2$ as large as $\al_s\cdot
\hscale^2$ -- in clear conflict with the conjecture stated above.

We will show now that this line of reasoning based on a naive physical
picture of the decay process is fallacious.
We will also explain what the loophole is in the
argument and that consistent calculations yield the correct result even
in such a simple ansatz.

The corrections to $\aver{M_{hadr}^2}$ that is generated by the
diagram in Fig.2  are given by multiplying the expression for the diagram
with $M\ind{hadr}^2=(p_B-q)^2=M_B^2(2x-1+Q^2)$.
This factor is analytical; therefore the integral of
$W(2x-1+Q^2)$ over $x$ can be evaluated via the usual
manipulations like deforming the integration path to the contour
$\be$ in Fig.3. In fact this factor
merely cancels out the pole associated with the gluon propagator. Again
the result has a regular limit when the spectator momenta vanish. In this
limit one arrives at the integration of a function that contains a double pole
at $x=(1-Q^2)/2$. The analysis then proceeds exactly like
in Section 2 and yields a regular four fermion amplitude
similar to the original one. Its matrix element then scales like $1/m_b^3$
and therefore it leads to corrections to
$\aver{M\ind{hadr}^2}$ only of
order $\hscale^3/m_b$ -- as conjectured. The essential
difference as compared to the total
probability is that the integration over $Q^2$ has now the form
$dQ^2\,(1-Q^2)$: the main contribution thus comes from high momenta
hadronic states. The correction is therefore given by the matrix
element of an operator different from the one that expresses the WA
contribution to the total width.

We have thus shown that WA does {\em not} induce corrections of
order $\hscale^2$ to $M\ind{hadr}^2$ after all:
they are suppressed at least by one
extra power of $1/m_b$. This result leads immediately to the question of
what could have gone wrong with the simple argument sketched above that
was based on a nonrelativistic picture of the $B$ meson?

The short answer is that the contributions from the two cuts across
the $u\bar u$ pair in the quark diagram become more
singular in the nonrelativistic limit and that
they have both
signs. Consider for example this forward scattering amplitude for
$b \bar u \ra b \bar u$ with the momentum of the spectator $\bar u$
being identical in the initial and final state: the left hand and the
right hand cut taken separately both become infinite. This singularity
is -- most conveniently and naturally --
regulated by assigning slightly different momenta
to the initial and final antiquarks; both cut contributions are
then finite though by themselves enhanced by the small difference in the
spacelike momenta of the initial and final $\bar u$
that appears in the denominator. It is easy to see that this
enhancement which reflects the original singularity disappears from
the sum of the two terms when one calculates the total decay probability.
Such a cancellation has to be expected on general grounds:
for the singularity reflects the emergence of a strong phase due to
one-gluon rescattering which does not affect the width directly.
On the other hand the singularity becomes
relevant in yielding a finite contribution to the invariant mass!
For the difference in the spectator momenta produces a corresponding
difference in the invariant mass for the two cuts; its small size is
compensated for by the nonrelativistic enhancements of the two
separate cut contributions. Thus we obtain a
contribution to the invariant mass even in the limit of
identical velocities
of the initial and final $\bar u$ antiquark that is
finite and positive and thus compensates for the negative contribution
obtained above!

It is instructive to analyze this problem from a more general
perspective as well because the apparent paradox can be formulated in
very general terms without referring to a nonrelativistic description.
The probability for the decay of a b quark into a final state
containing an on-shell gluon without a $u \bar u$ pair is
of course positive and scales like $1/m_b$. As the WA
correction to the total
decay width is of order $1/m_b^3$, the probability for b
quark decays to final
states with a $u \bar u$ pair has to scale like $1/m_b$ and be negative.
The typical mass square of such states is of order
$\hscale m_b$; therefore they lead to a large negative correction to
$\langle M^2\ind{hadr}\rangle$.
This paradox has an interesting solution that can be formulated
in obvious analogy to problems in ordinary quantum
mechanics when one considers for simplicity a system with a discrete spectrum
of
states rather than a continuous one. For calculating the corrections
to $\langle M\ind{hadr}^2 \rangle$ one
has to consider not only the corrections to the
probability of the decays to $u \bar u$ states but also to
include the shift in their energy levels. Because the massless
gluon has an energy less than the energy
of the $q\bar q$ pair, its exchange {\em raises} the invariant mass of the
latter, and therefore provides the missing positive contribution to
$M\ind{hadr}^2$.

This standard quantum mechanical effect
is of course automatically contained in a computation based on Feynman
diagrams. What we have demonstrated above is that the necessary cancellations
arise also
as long as one deals carefully with the
singularities that emerge in a nonrelativistic treatment of quark diagrams.
There is a  general lesson to be learnt from it:
{\em a priori} one has to be very careful
in drawing conclusions
about small nonperturbative corrections appearing on top of
large free parton amplitudes. More specifically our analysis
in this section shows once again that some effects of WA can
emerge in rather delicate ways. A failure to
treat properly all possible
contributions where identical $q\bar q$ pairs are present
can easily lead to a dramatic
overestimate of the impact of WA.
To cite but one example: WA can certainly induce
scattering phases in some exclusive modes and thus affect
their transition rates significantly; yet from such an observation
alone one cannot infer that WA is numerically important
in inclusive $B$ and $D$ decays.

To conclude this section we add the comment that an
analysis similar to the one outlined here can be
applied also to
interference effects in semileptonic decays of heavy flavour baryons.
PI contributes first on the level $1/m_b^3$ and thus can cause a shift
in $M\ind{hadr}^2$
at most of order $\hscale^3/m_b$.

 \section {Summary and Outlook}

When the first data on D meson lifetimes showed a large difference
between $\tau (D^+)$ and $\tau (D^0)$ it was immediately realized
that WA can shorten $\tau (D^0)$ considerably --  if its inherent
helicity suppression can be overcome. In particular it was suggested
that the presence of gluons could enhance WA contributions.
Ref.~\cite{BSS} considered the perturbative emission of energetic gluons.
Yet our analysis in ref.~\cite{BU} showed that perturbative gluon radiation
can vitiate helicity suppression only at a small numerical level of order
$\alpha_S(m_b^2)$.

Another attempt~\cite{FM} invoked nonspectator contributions in the
form of soft gluons in the wavefunction of heavy flavour hadrons.
In the present paper we have discussed in considerable detail the impact
of weak annihilation on
weak decays of heavy flavour hadrons with the main emphasis on
semileptonic $b\ra
u$ decays, since those are easier to treat theoretically.
The results we found are in broad qualitative agreement with the ansatz of
ref.~\cite {FM} -- as long as it makes sense at all to speculate about
nonperturbative dynamics in terms of individual gluons.
Gluons in the
wavefunctions could -- in an oversimplified picture --
generate the nontrivial intermediate states
saturating the local four fermion operator in eq.(6):
for the quark current then annihilates only the constituent quarks in
the wavefunction leaving extra quanta in the final state. Of course the
agreement between our analysis and that of ref.~\cite{FM} is only of a
very rough qualitative nature. Among other differences we have shown that
the relative weight of such nonspectator contributions scales like
$1/m_Q^3$ and not like $1/m_Q$ or $1/m_Q^2$ as conjectured in
ref.~\cite{FM}.

More specifically we have shown that nonperturbative effects could generate a
significant difference in $\Gamma (B^0\ra l\nu + X_u)$ vs.
$\Gamma (B^-\ra l\nu + X_u)$
that
is {\em not} suppressed by the mass of the lepton and {\em a priori}
could be 6 times as large (for $g \simeq 1/3$) as
$\Gamma (B\ra \tau\nu_\tau)$ for
both electron and muon decays; this would represent a significant
contribution, namely $\sim 20\%$
of the total $b\ra ul \nu$ width. Furthermore this difference
is concentrated in a narrow slice -- with a width $\hscale/2$ --
just below the endpoint of the lepton energy spectrum;
for real B decays it is the region starting around $2.3\GeV$.
Therefore if one studies only the region {\em above} the $b\ra c$ endpoint the
relative weight of WA in semileptonic $B^-$ decays
gets enhanced and could actually become large.
This additional
contribution in $B^\pm$ decays can
significantly affect the value of $|V_{ub}|$ that up to now has been
derived from
semileptonic spectra averaged over charged and neutral $B$'s.

The exact numerical size of the WA contributions to the lepton
spectrum depends on the magnitude of two matrix elements; one
involves colour singlet and the other colour octet operators.
In the endpoint region, the domain of nonperturbative dynamics,
the colour singlet operator dominates. There exists also a
``logarithmic'' tail extending below the endpoint region where
a perturbative treatment applies; colour octet operators play a larger
role there.
It should be noted that numerically annihilation effects could be
larger than the flavour independent effects identified in
ref.~\cite{BSUV} although the latter are formally leading in $1/m_b$
(see ref.~\cite{BUV}).

Obviously inclusive charm decays can be
analyzed in a quite analogous fashion
since the heavy quark symmetry ensures that the same four fermion matrix
elements enter in the decay rates of charm hadrons;
the only difference arises due to hybrid renormalization which mixes
the colour singlet and octet operators and which can be calculated
perturbatively. Of course the nonperturbative corrections are much larger
in $D^+\ra l\nu X$ vs. $D^0\ra l\nu X$ than in $B^-\ra l\nu X$ vs.
$B^0\ra l\nu X$ decays
due to $1/m_c > 1/m_b$ and also $|V(cd)/V(cs)| > |V(ub)/V(cb)|$;
even more prominent effects are expected in $D_s\ra l\nu X$ decays.
This makes such an analysis of charm decays more challenging as well as
more intriguing. We will report on it in a future paper~\cite{BUR}.

Our results are rather qualitative or at best semi-quantitative in nature.
This is not surprising since we are dealing with nonperturbative
corrections in general and the size of nonfactorizable matrix elements in
particular. At present we have little theoretical guidance in
evaluating these features in a numerically reliable fashion: folklore has it
that the nonfactorizable parts in matrix elements are suppressed by the
number of colours; thus one expects $g\ind{singl}$ to be less than unity in
the real world. At the same time however the suppression of $g\ind{oct}$ may
appear to be softer. On the other hand QCD sum rules tend to suggest that
nonfactorizable contributions to the $\de_{\mu\nu}$ term are in general
smallish in heavy flavour mesons~\cite{VLB}; yet further studies are
obviously required before reliable conclusions can be drawn.

Semileptonic $\Lambda _b$ decays provide an intriguing lab to study
effects due to both WS and PI. They can be analyzed
in close analogy to the discussion given in this paper.
For example PI affects (KM suppressed) semileptonic
$\Lambda _b$ decays and it does so mainly in the endpoint region --
as it is the case for WA in $B$ meson decays.
Since WS is {\em not} generally
helicity suppressed its relative impact on
preasymptotic corrections could be sizeable.
Yet for that very reason one cannot expect nonperturbative
low energy scale physics
to play a prominent and easily
identifiable role in $\Lambda _b$ decays.
Furthermore here we do not have an unambigous yardstick as
we had it in the comparison
between charged and neutral $B$ mesons. All of this will make it
considerably more difficult to separate out various preasymptotic
effects.

$B_c$ mesons constitute a very intriguing
dynamical system. With the charm quark mass providing an infrared
cut-off $B_c$ decays can be treated in perturbation theory. WA is
not KM suppressed here and the hadronic matrix element that determines
the strength of WA can be calculated both in its factorizable and
nonfactorizable part. Furthermore final states from WA that contain a
$c\bar c$ pair should be experimentally separable from those with a gluon
(or gluons) instead in the final state. For the latter one can then
invoke the nonrelativistic calculation of ref.~\cite {BSS} with its
enhancement factor $(m_b/m_c)^2$. These calculations are not trivial,
but they appear to be doable in a straightforward way. Obtaining
a sample of $B_c$ decays sufficiently large to make the studies referred
to above feasible represents a stiff experimental challenge, yet it might
not be beyond our reach.

There exist many cross references between a theoretical understanding of
semileptonic and nonleptonic heavy flavour decays. More specifically the
preasymptotic corrections that we have been discussing in this note
appear already in the KM allowed $b\ra c$ transitions, namely WA in
$B_d$ and $B_s$, PI in $B^{\pm}$ decays. We then infer that the
impact of WA and PI on total nonleptonic decay rates can be expressed
in terms of matrix elements of two four fermion operators that differ
solely in the way colour flows through them. We also conjecture that the
weight of WA in the total width is determined mainly by the same
nonfactorizable piece coming from nonperturbative low energy dynamics
that can be extracted from a detailled study of semileptonic $B$ decays.
A considerably larger correction in nonleptonic decays is expected from
PI which appears already in factorizable contributions
(see refs.~\cite{VSlog,BU} for details). For this reason
we view our earlier estimate as reliable -- unless an
unexpectedly large difference in the lepton spectra in $B^0$ vs.
$B^-$ decays forces us to revise the common wisdom about the
numerical validity of the factorization approximation in the real
world; in that case one had to reconsider the hierarchy of the
preasymptotic corrections outlined above.

\vspace*{0.5cm}

{\bf ACKNOWLEDGEMENTS:} \hspace{.4em} We gratefully acknowledges
illuminating discussions and exchanges on the subject of this paper with
several colleagues, and in particular with V.Braun, Yu.Dokshitzer,
A.Vainshtein and A.Johansen. This work was supported in part by the National
Science Foundation under grant number PHY 92-13313.

\vspace*{1cm}
{\LARGE{\bf Figure Captions}}
\vspace*{0.5cm} \\
{\Large{\bf Fig.1 }} Parton diagrams for the simplest WA processes:\\
{\bf a)} ``Parton annihilation'' at tree level.\\
{\bf b)} Conventional gluon annihilation.\\
{\bf c)} Interference in the ``quasifree'' mode $b\ra u +
\bar q q'\:(l\nu_l)$ induced by the gluon WA amplitude.
\vspace*{.3cm}\\
{\Large {\bf Fig.2 }} Lowest order perturbative diagram for the WA-induced
correction to correlators $W$ in eq.(1).
\vspace*{.3cm}\\
{\Large {\bf Fig.3 }} The complex $x$ plane. Crosses denote the poles
representing singularities of the one gluon annihilation amplitude. The
heavy line shows the cut of the Jacobian in eq.(2).
\vspace*{.3cm}\\
{\Large {\bf Fig.4 }} Shape of the difference in the lepton spectra
for $B^{\pm}\ra X_u l\nu$ vs. $B^0\ra X_u l\nu$ decays corresponding to
$g\ind{oct}/g\ind{singl}=-1/2$ (solid line) and $g\ind{oct}/g\ind{singl}=1/3$
(dashed line).  Below region A ($E_1 < 2.6\GeV$) the curves describe the
spectrum given by
eq.(15) with $\Lambda\ind{QCD}=180\MeV$. The thick vertical line at
$E_l=M_B/2$ represents the monochromatic line from the chirality
suppressed two body mode $B^{\pm}\ra l\nu$.


\begin{thebibliography}{17}

\bibitem{we}
I.I.Bigi, N.G.Uraltsev, A.Vainshtein, {\it Phys.Lett.} {\bf 293B}
(1992) 430;\\
B.Blok, M.Shifman, {\it Nucl.Phys.}{\bf B399} (1993)441, 459; \\
I.I.Bigi, B.Blok, M.Shifman, N.G.Uraltsev, A.Vainshtein,
{\bf preprint} UND-HEP-92-BIG07, TPI-MINN-92/67-T.

\bibitem{VS}
M.A.Shifman, M.B.Voloshin, {\it Sov.Journ.Nucl.Phys.} {\bf 41} (1985) 120.

\bibitem{VSlog}
M.A.Shifman, M.B.Voloshin,  {\it Sov.Phys.ZhETF} {\bf 64} (1986) 698; {\it
Sov.Journ.Nucl.\\Phys.} {\bf 45} (1987) 292.

\bibitem{BU}
I.I.Bigi, N.G.Uraltsev, {\it Phys.Lett.} {\bf 280B} (1992) 120;
more details can actually be found in the preprint version, UND-HEP-92-BIG02.

\bibitem {CGG}
J.Chay, H.Georgi and B.Grinstein, {\em Phys.Lett.}~{\bf B247} (1990) 399.

\bibitem{BUV}
I.I.Bigi, N.G.Uraltsev, A.Vainshtein, {\it Phys.Lett.} {\bf 293B} (1992)
430.

\bibitem{BUR}
I.I.Bigi, N.G.Uraltsev, {\bf preprint} UNDHEP-BIG03 (1993).

\bibitem{BSS}
M.Bander,D.Silverman, A.Soni, {\it Phys.Rev.Lett.}~{\bf 44}~(1980)~7,~962(E).

\bibitem{BSUV}
I.I.Bigi, M.Shifman, N.G.Uraltsev, A.Vainshtein, {\em Phys.Rev.Lett.}
{\bf 71}~(1993)~496.

\bibitem{ACM}
G.Altarelli et al., {\em Nucl.Phys.} {\bf B208} (1982) 365.

\bibitem{BSUV2}
I.I.Bigi, M.Shifman, N.G.Uraltsev, A.Vainshtein, {\it in preparation}.

\bibitem{BKSV}
B.Blok, L.Koyrakh, M.Shifman, A.Vainshtein, {\bf preprint}
TPI-MINN-93/33-T.

\bibitem{FM}
H.Fritzsch, P.Minkowski, {\it Phys.Lett.} {\bf 90B} (1980) 455.

\bibitem{VLB}
V.Braun, private communication.


\end{thebibliography}
\end{document}